\documentclass[10pt,conference]{IEEEtran}
\IEEEoverridecommandlockouts

\usepackage{cite}
\usepackage{amsmath,amssymb,amsfonts}
\usepackage{algorithmic}
\usepackage{graphicx}
\usepackage{textcomp}
\usepackage{xcolor}
\def\BibTeX{{\rm B\kern-.05em{\sc i\kern-.025em b}\kern-.08em
    T\kern-.1667em\lower.7ex\hbox{E}\kern-.125emX}}

\usepackage{bm}
\usepackage{tabularx}
\usepackage{mathtools}
\usepackage{multirow}
\usepackage{subfig}
\usepackage{hyperref}

\usepackage{array}
\newcolumntype{M}[1]{>{\centering\arraybackslash}m{#1}}
\newcolumntype{P}[1]{>{\centering\arraybackslash}p{#1}}
\newcolumntype{L}[1]{>{\raggedright\arraybackslash}m{#1}}

\newcommand{\Dcal}{\mathcal{D}}
\newcommand{\Ecal}{\mathcal{E}}

\newcommand{\Ncal}{\mathcal{N}}

\newcommand{\Qcal}{\mathcal{Q}}

\newcommand{\Scal}{\mathcal{S}}

\newcommand{\Vcalsat}{\mathcal{V}_{\text{S}}}
\newcommand{\Vcalgw}{\mathcal{V}_{\text{G}}}

\newcommand{\Ecalul}{\mathcal{E}_{\text{UL}}}
\newcommand{\Ecalisl}{\mathcal{E}_{\text{ISL}}}
\newcommand{\Ecalfl}{\mathcal{E}_{\text{FL}}}

\newcommand{\Ecalbar}{\bar{\mathcal{E}}}

\newcommand{\nn}{\nonumber}

\begin{document}

\title{QoS Identifier and Slice Mapping in 5G and Non-Terrestrial Network Interconnected Systems}

\author{
\IEEEauthorblockN{Yuma Abe, Mariko Sekiguchi, and Amane Miura}
\IEEEauthorblockA{Space Communication Systems Laboratory, Wireless Networks Research Center, Network Research Institute, \\ National Institute of Information and Communications Technology (NICT), Tokyo, Japan, \\ \emph{\{yuma.abe, sekiguchi, amane\}@nict.go.jp}}
}

\maketitle

\begin{abstract}
The interconnection of 5G and non-terrestrial networks (NTNs) has been actively studied to expand connectivity beyond conventional terrestrial infrastructure.
In the 3GPP standardization of 5G systems, the 5G Quality of Service (QoS) Identifier (5QI) is defined to characterize the QoS requirements of different traffic requirements.
However, it falls short in capturing the diverse latency, capacity, and reliability profiles of NTN environments, particularly when NTNs are used as backhaul.
Furthermore, it is difficult to manage individual traffic flows and perform efficient resource allocation and routing when a large number of 5G traffic flows are present in NTN systems.
To address these challenges, we propose an optimization framework that enhances QoS handling by introducing an NTN QoS Identifier (NQI) and grouping 5G traffic into NTN slices based on similar requirements.
This enables unified resource control and routing for a large number of 5G flows in NTN systems.
In this paper, we present the detailed procedure of the proposed framework, which consists of 5QI to NQI mapping, NTN traffic to NTN slice mapping, and slice-level flow and routing optimization.
We evaluate the framework by comparing multiple mapping schemes through numerical simulations and analyze their impact on overall network performance.
\end{abstract}


\section{Introduction}

The interconnection of 5G and non-terrestrial networks (NTNs) has attracted attention as a way to extend connectivity beyond conventional terrestrial infrastructure~\cite{Bakambekova_OJCS24}.
NTNs include various platforms, such as geostationary (GEO) and low Earth orbit (LEO) satellites, high-altitude platform stations (HAPS), and drones.
To support this interconnection, the Third Generation Partnership Project (3GPP) has initiated standardization efforts, and specifications are under development.

Several frameworks have been proposed for interconnecting 5G and NTNs.
Kak and Akyildiz~\cite{Kak_TNSM22} proposed an automatic network slicing framework for CubeSat-based space-ground integrated networks, formulating the slicing as a mixed integer nonlinear program with a topology design and online segment routing to reduce service level agreement (SLA) violations.
Drif et al.\cite{Drif_LCN21} introduced an end-to-end slicing architecture that ensures slice and QoS continuity across 5G and satellite domains through flow translation and control plane extensions.
Liu et al.\cite{Liu_TNSM24} developed a Quality of Service (QoS)-aware framework that integrates user plane functions (UPFs) into LEO satellite networks, combining static user-to-UPF assignment with dynamic migration to reduce latency and energy consumption.

\begin{table}[t]
  \centering
  \caption{Selected 5QIs defined in 3GPP TS 23.501~\cite{3GPP_TS23.501}, focusing on Resource Type, Packet Delay Budget, and example services.}
    \scalebox{0.85}{
    \begin{minipage}{\textwidth}
    \begin{tabular}{cccL{5.2cm}} \hline 
    5QI & Resource & Packet Delay & \multirow{2}{*}{Example Services} \\ 
    Value & Type & Budget & \\ \hline \hline
    1 & GBR & 100~ms & Conversational Voice \\
    2 & GBR & 150~ms & Conversational Video (Live Streaming) \\
    3 & GBR & 50~ms & Real Time Gaming, V2X messages\\
    4 & GBR & 300~ms & Non-Conversational Video (Buffered Streaming)\\
    65 & GBR & 75~ms & Mission Critical user plane Push To Talk voice\\
    71 & GBR & 150~ms & ``Live'' Uplink Streaming \\
    \hline
    5 & Non-GBR & 100~ms & IMS Signalling \\
    6 & Non-GBR & 300~ms & Video (Buffered Streaming), TCP-based \\
    10 & Non-GBR & 1100~ms & Video (Buffered Streaming), TCP-based, and any service over satellite access \\
    70 & Non-GBR & 200~ms & Mission Critical Data \\
    79 & Non-GBR & 50~ms & V2X messages \\
    80 & Non-GBR & 10~ms & Low Latency eMBB applications, Augmented Reality \\
    \hline
    \end{tabular}
    \vspace{1mm}
    \\
    V2X: Vehicle-to-Everything, IMS: IP Multimedia Subsystem, \\ IP: Internet Protocol, TCP: Transmission Control Protocol, \\ eMBB: enhanced Mobile Broadband
    \end{minipage}
    }
\label{table:5QI}
\end{table}

5G QoS Identifier (5QI) values are defined in the 3GPP specification to describe the QoS requirements of traffic for frequently used services in 5G systems~\cite{3GPP_TS23.501}.
These values specify standardized QoS characteristics for the required packet forwarding treatment between User Equipment (UE) and the UPF of 5G Core.
We describe examples of the standardized 5QIs in Table~\ref{table:5QI}.
By definition, each 5QI value is mapped to QoS characteristics, e.g., Resource Type, Default Priority Level, Packet Delay Budget, and Packet Error Rate, and their corresponding example services are also specified.
The Resource Type includes Guaranteed Bit Rate (GBR) and Non-GBR.
The Packet Delay Budget (PDB) defines an upper bound of latency in which a packet may be delayed between the UE and the UPF. 
For example, 5QI 1 and 6 are used for ``Conversational Voice'' (GBR, PDB: 100~ms) and ``Video (Buffered Streaming)'' (Non-GBR, PDB: 300~ms), respectively.
The first NTN-related 5QI was introduced as 5QI 10 with a PDB of 1,100~ms in 3GPP Release 17.
These identifiers enable 5G systems to recognize traffic characteristics and allocate resources accordingly.

In configurations where 5G systems use NTNs as backhaul, 5G UEs connect to 5G base stations (gNBs), and their traffic is routed to designated data networks (DNs) via the gNB, NTN links, NTN ground stations, and the 5G Core's UPF.
To support this, gNBs must interface with NTN nodes such as satellites.
The latency of NTN links varies significantly, ranging from 30~ms (LEO) to 500~ms (GEO), which makes it difficult for a single 5QI value like 5QI 10 (PDB: 1,100~ms) to represent diverse traffic requirements in NTN systems accurately.
Furthermore, it is difficult to manage each unit of traffic and calculate resource allocation and routing, especially when a large number of 5G traffic is present in the systems.
To address this, we introduce the \textit{NTN QoS Identifier (NQI)} to represent traffic QoS in NTNs, and the concept of an \textit{NTN slice}, which groups traffic based on QoS classifications~\cite{Sekiguchi_IAC23}.
This framework enables unified handling of 5G and NTN traffic and facilitates efficient control through slice-based traffic management.
Similar concerns have been raised in the literature and European research initiatives.
The SaT5G project noted the granularity gap between 5QI and satellite service classes and proposed mapping mechanisms to cope with NTN link variability~\cite{SaT5G_D3.2_18}.
Along the same lines, Drif et al.\cite{Drif_IJSCN20} emphasized that a single mapping between standardized 5QI values and satellite QoS classes is insufficient to guarantee end-to-end performance across terrestrial and satellite domains.
These studies motivate the need for a new QoS abstraction tailored for NTNs, which we capture through the proposed NQI and NTN slice concepts.

In this paper, we propose an optimization framework for end-to-end traffic flow in a 5G–NTN integrated system.
The framework consists of (i) 5QI to NQI mapping, (ii) NTN traffic to NTN slice mapping, and (iii) slice-level flow allocation and routing optimization.
We evaluate the proposed framework by comparing multiple mapping schemes and analyzing their impact on NTN performance through numerical simulations.

\section{System Model}

In this section, we describe a system model regarding network, user, and traffic aspects in the 5G-NTN interconnected systems.

\subsection{NTN Configuration}

We model satellites and ground stations in the NTN as $\Vcalsat$ and $\Vcalgw$, respectively.
Regarding links in the NTN, we consider three types: (i) user links, connecting a user and a satellite; (ii) inter-satellite links, connecting satellites; and (iii) feeder links, connecting a satellite and a ground station.
The corresponding link sets are denoted by $\Ecalul$, $\Ecalisl$, and $\Ecalfl$, respectively, with the complete set of links given by $\Ecal = \Ecalul \cup \Ecalisl \cup \Ecalfl$.
The availability of these links varies dynamically over time, depending on link status, which may change due to satellite movement, weather conditions, or potential failures.

\subsection{User and Traffic Configuration}

We define ``5G traffic'' as traffic generated by a 5G UE with a 5QI and ``NTN traffic'' as traffic generated by NTN users—including aircraft, ships, and also gNBs in this paper—with an NQI.
In this paper, we assume each 5G UE can generate multiple traffic flows, and we focus on traffic flowing from 5G UEs to DN via NTN ground stations.
Here, we assume the presence of $N_{g}$ gNBs, each of which is connected to multiple users, and these users generate a total of $N_{u,j}$ 5G traffic flows from the $j$-th gNB.
The sets of gNB and 5G traffic indices are denoted by $\Ncal_{g}=\{1,2,\ldots,N_{g}\}$ and $\Ncal_{u,j}=\{1,2,\ldots,N_{u,j}\}$, respectively.
The set of destination candidates is denoted by $\Dcal = \{d_{1}, d_{2}, \ldots, d_{N_{d}}\}$, where $N_{d}$ is the total number of destination candidates.

The sets of 5QIs and NQIs are denoted by $\Qcal_{f}=\{q_{f}\}$ and $\Qcal_{n}=\{q_{n}\}$, where $q_{f}$ and $q_{n}$ represent the 5QI and NQI, respectively.
The PDB associated with each QoS identifier $q$ (either 5QI or NQI) is denoted by $\bar{\ell}(q)$.
The $i$-th 5G and NTN traffic flows are represented by tuples $t_{f,i}(q_{f,i}, r_{f,i}, d_{i})$ and $t_{n,i}(q_{n,i}, r_{n,i}, d_{i})$, respectively, where $r_{f,i}\geq 0$ and $r_{n,i}\geq 0$ denote the respective flow demand, and $d_{i}\in\Dcal$ denote the destination.

\section{Optimization Framework for End-to-End Traffic Flow}
\label{section:optimization_framework}

\begin{figure}[tb]
	\centering
	\includegraphics[width=\columnwidth]{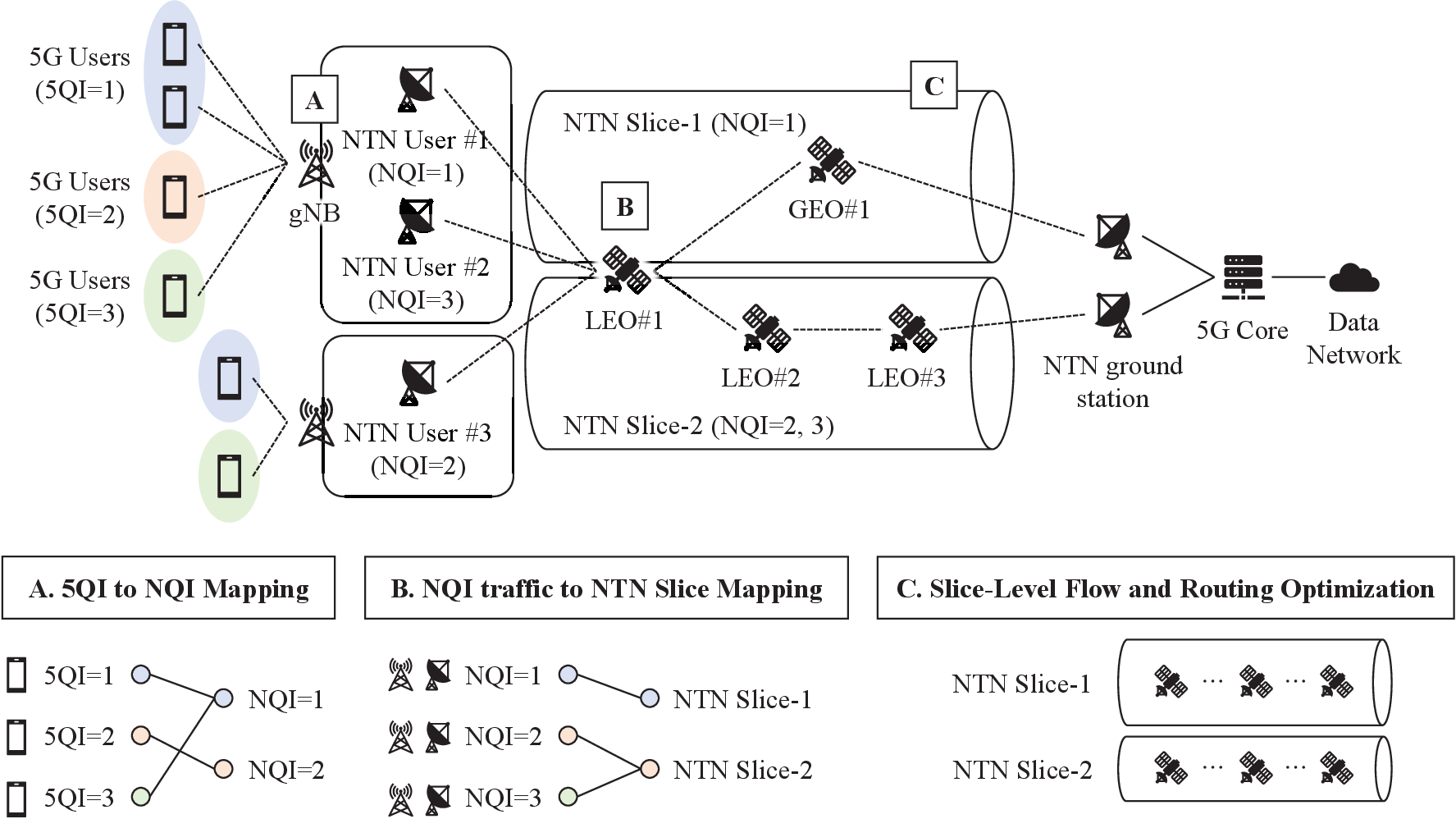}
	\caption{Proposed optimization framework for 5G-NTN interconnected systems. The equations shown in this figure will be introduced in Section~\ref{section:optimization_framework}.}
\label{fig:proposed_framework}
\end{figure}

We propose an optimization framework for end-to-end traffic flow in 5G-NTN interconnected systems, as shown in Fig.~\ref{fig:proposed_framework}.
This framework is divided into three phases: ``5QI to NQI mapping,'' ``NTN traffic to NTN slice mapping,'' and ``Slice-level routing and resource control''.
The remainder of this section provides a detailed explanation of each phase.

\subsection{5QI to NQI Mapping}

In the network, there are two types of traffic: 5G traffic and NTN traffic.
To correctly handle traffic within the NTN, 5G traffic is transformed into NTN traffic.

When a 5G UE is connected to a gNB, the 5QI of each traffic from the UE is mapped to an NQI at the corresponding gNB.
We define a discrete mapping function from 5QI to NQI as $q_{n} = f_{j}(q_{f})$, where $q_{n}\in\Qcal_{n}$ and $q_{f}\in\Qcal_{f}$ and $f_{j}$ is the mapping function used by the $j$-th gNB.
Note that the function $f_{j}$ can differ between gNBs.
In addition, at each gNB, traffic flows with the same NQI are aggregated and treated as a single NTN traffic flow.
Therefore, in the following discussion, gNBs are regarded as a type of NTN user.
The flow demand of the NTN traffic is calculated by that of the aggregated 5G traffic, represented by $r_{n,i}=\sum_{i\in\Ncal_{q_{n},j}}r_{f,i}$, where the set $\Ncal_{q_{n},j}\subseteq\Ncal_{u,j}$ represents the traffic flows assigned the same NQI $q_{n}$ at the $j$-th gNB.

The mapping between 5QI and NQI significantly affects the flow allocation and routing performance for each gNB.
For example, if traffic with a small PDB (as specified by its 5QI) is mapped to an NQI with a larger PDB, the assigned route may fail to meet the original latency requirement.
Conversely, if the small PDB constraint is strictly preserved during NTN route selection, it may become impossible to find a feasible route within the NTN.

\subsection{NTN Traffic to NTN Slice Mapping}
\label{sec:nqi_slice_mapping}

We define an ``NTN slice'' as a set of NTN traffic flows that share the same destination and one or more specified NQIs.
In this phase, traffic flows with designated NQIs are aggregated and mapped to an NTN slice at each satellite.
Note that an NTN slice is defined between a satellite and a DN.
Here, we call the first satellite attached for each NTN user a ``slice-edge satellite'' and define a subset of slice-edge satellites as $\Vcalsat'\subseteq\Vcalsat$.
The slice-edge satellite for each user is determined using appropriate methods or algorithms.

The $\ell$-th slice-edge satellite has a discrete mapping function from the NTN traffic to the NTN slice index, denoted by $j = g_{\ell}(t_{n})$.
This mapping function can differ between satellites.
By using this function, each NTN traffic flow 
$t_{n,i}(q_{n,i}, r_{n,i}, d_{i})$ is then mapped to the $j$-th NTN slice defined as 
\begin{align}
    \Scal_{\ell,j} = \{t_{n,i}(q_{n,i}, r_{n,i}, d_{i}) \mid q_{n,i} \in \Qcal_{\ell,j}, d_{i} = d_{j} \in \Dcal \}. \nn
\end{align}
The set $\Qcal_{\ell,j}\subseteq\Qcal_{n}$, with $j \in \{1, 2, \ldots, N_{q,\ell}\}$, 
represents the NQI set mapped to the $j$-th slice, and $N_{q,\ell}$ is the number of such NQI groups.
The notation $d_{j}$ denotes the destination associated with the $j$-th NTN slice. 
The maximum number of NTN slices that can be created at the $\ell$-th satellite 
is $|\mathcal{D}| \times N_{q,\ell}$.

For the following formulation, we define the set of slices generated in the system as $\Scal = \{\Scal_{1}, \Scal_{2}, \ldots, \Scal_{N_{s}}\}$, and the corresponding index set as $\Ncal_{s} = \{1, 2, \ldots, N_{s}\}$, where $N_{s} = |\Scal|$.
The total requested traffic flow for the $i$-th slice is denoted by $r_{s,i}$, and it is calculated by
\begin{align}
    r_{s,i} = \sum_{j:t_{n,j} \in \Scal_{i}} r_{n,j}.
    \label{eq:slice_required_flow}
\end{align}

\subsection{Slice-Level Flow and Routing Optimization}
\label{sec:slice_optimization}

Finally, we determine the flow allocation and routing for the NTN slices defined so far.
To this end, we define an optimization problem for all the NTN slices in the network, considering the network's performance in terms of allocated flow and latency as key metrics to include in the cost function~\cite{Kak_TNSM22,Abe_VTC-Spring24}.

First, we define decision variables for an optimization problem.
We define a binary variable of link assignment for each slice, $x_{i,e}, \forall i\in\Ncal_{s}, e \in \Ecal$, as follows:
\begin{align}
    x_{i,e} =
    \begin{cases}
        1, & \text{if $i$-th slice is assigned to a link $e$,} \\
        0, & \text{otherwise.}
    \end{cases}
\nn
\end{align}
Here, a set of links assigned to $i$-th slice is represented by $\Ecalbar_{i} = \{e \mid x_{i,e} = 1\}\subseteq \Ecal$.
Additionally, we denote by $b_{i} \geq 0$ the amount of flow intended for the $i$-th slice, and by $f_{i,e} \geq 0$ the flow amount on link $e$ for the $i$-th slice.
These two parameters $b_{i}$ and $f_{i,e}$ are subject to the following relationship:
\begin{align}
    b_{i} = \sum_{e\in\Ecalbar_{i}} f_{i,e},~~\forall i\in\Ncal_{s}.
    \label{eq:slice_b_equal_sum_f}
\end{align}

The constraints to be satisfied in the optimization problem are summarized in Table~\ref{table:constraints}.
We define two functions, $u(e)$ and $v(e)$, to map a link $e$ to its egress and ingress nodes, respectively.
The notation $c_{e}$ is a corresponding capacity of a link $e$.
The detailed explanation of these constraints is omitted; please refer to~\cite{Abe_VTC-Spring24} for more detailed information.

\begin{table*}[tb]
\vspace*{0.2cm}
    \centering
    \caption{List of constraints and their objectives}
    \scalebox{1}{
        \begin{tabular}{p{3.5cm}lp{6cm}} \hline
            Name & Constraint & Objective/Explanation \\ \hline \hline
            (i) Feasibility constraint
            & $f_{i,e} \leq c_{e} x_{i,e},~~\forall i\in\Ncal_{s}, e\in\Ecalisl\cup\Ecalfl$
            & Ensure that traffic flows only through active links.
            \\
            (ii) Capacity constraint
            & $\sum_{\substack{i\in\Ncal_{s}}} f_{i,e} \leq c_{e}~\forall e \in \Ecalisl\cup\Ecalfl$
            & Ensure that the total flow on each link does not exceed its capacity.
            \\
            (iii) Flow conservation constraint at slice-edge satellites
            & $\sum_{\substack{j = v(e), \\ e\in\Ecalisl}} f_{i,e} + b_{i} = \sum_{\substack{j = u(e), \\ e\in\Ecalisl\cup\Ecalfl}} f_{i,e},~~\forall i\in\Ncal_{s}, j\in\Vcalsat'$
            & Ensure that flow is conserved at each slice-edge satellite, which handles the NTN slice's $b_{i}$. \\
            \\
            (iv) Flow conservation constraint at other satellites
            & $\sum_{\substack{j = v(e), \\ e\in\Ecalisl}} f_{i,e} = \sum_{\substack{j = u(e), \\ e\in\Ecalisl\cup\Ecalfl}} f_{i,e},~~\forall i\in\Ncal_{s}, j\in\Vcalsat$
            & Ensure that flow is conserved at each relay satellite. \\
            \\
            (v) Destination node flow constraint
            & $\sum_{\substack{j = v(e),\\ e \in \Ecalfl}} f_{i,e} = b_{i}, \sum_{\substack{j = u(e),\\ e \in \Ecalfl}} f_{i,e} = 0,~~\forall i\in\Ncal_{s}, j\in\Vcalgw$
            & Ensure that the destination node receives flow $b_{i}$ and does not transmit any further traffic. \\
            \\
            (vi) Total number flow conservation constraint
            & $\sum_{\substack{j = v(e), \\ e\in\Ecalisl}} x_{i,e} = \sum_{\substack{j = u(e), \\ e\in\Ecalisl\cup\Ecalfl}} x_{i,e},~~\forall i\in\Ncal_{s}, j\in\Vcalsat$
            & Ensure that link usage is balanced at relay nodes for each slice.
            \\
            (vii) No-loop constraint
            & $\sum_{\substack{j = v(e), \\ e\in\Ecalisl}} x_{i,e} \leq 1, \sum_{\substack{j = u(e), \\ e\in\Ecalisl}} x_{i,e} \leq 1,~~\forall i\in\Ncal_{s}, j\in\Vcalsat$
            & Ensure that no routing loop is formed in any selected link.
            \\
            (viii) Connection number constraint in feeder links
            & $\sum_{\substack{d_i = v(e) \\ e \in \Ecalfl}} x_{i,e} \leq 1,~~\forall i\in\Ncal_{s}$
            & Ensure that each slice uses at most one feeder link to reach its destination.
            \\
            \hline
        \end{tabular}
    }
\label{table:constraints}
\end{table*}

The 5G–NTN interconnected systems attempt to allocate flow to satisfy the slice's flow request as much as possible and assign end-to-end links to meet the slice's allowable latency, i.e., PDB, to the greatest extent possible.
To quantify this, we define the flow gap and latency gap as follows.
First, the flow gap is defined as the difference between the flow requirement and the allocated flow for each slice, as follows:
\begin{align}
    s_{f,i} = \text{max}(0, r_{s,i} - b_{i}),
    \label{eq:satisfaction_flow}
\end{align}
where $r_{s,i}$ represents the total requested flow of the $i$-th slice defined in Eq.~(\ref{eq:slice_required_flow}).
This metric indicates that the flow gap decreases as the allocated flow approaches the requirement, and becomes zero once the allocation meets or exceeds the demand.

Next, let $\ell_{i}$ denote the latency of the $i$-th slice, calculated as:
\begin{align}
    \ell_{i} = \sum_{e\in\Ecalbar_{i}} \bar{\ell}_{e} x_{i,e},~\forall i\in\Ncal_{s}, \nn
\end{align}
where $\bar{\ell}_{e}$ denote the propagation latency of link $e$.
Using this expression, we define the latency gap as:
\begin{align}
    s_{\ell,i} = \text{max}(0, \ell_{i} - \bar{\ell}(q_{n,i})).
    \label{eq:satisfaction_latency}
\end{align}
Here, $\bar{\ell}(q_{n,i})$ corresponds to the PDB value associated with the NQI of the $i$-th slice.
The latency gap takes a positive value when $\ell_{i}$ exceeds $\bar{\ell}(q_{n,i})$, and becomes zero when $\ell_{i}$ is less than or equal to $\bar{\ell}(q_{n,i})$.

We then define a cost function that considers both flow allocation and latency as follows:
\begin{align}
    J = \dfrac{w_{f}}{N_{s}F} \sum_{i\in\Ncal_{s}} s_{f,i} + \dfrac{w_{\ell}}{N_{s}L} \sum_{i\in\Ncal_{s}} s_{\ell,i},
\label{eq:cost}
\end{align}
where $w_{f} \geq 0$ and $w_{\ell} \geq 0$ are the weights for each term, chosen such that $w_{f} + w_{\ell} = 1$, and $F$ and $L$ are normalization constants for flow and latency, respectively, chosen to ensure that $J_{f}$ and $J_{\ell}$ fall within the range $[0,1]$.
These metrics can be easily linearized by introducing slack variables, as discussed in~\cite{Boyd_04}.

We formulate a joint flow and routing optimization problem for all slices as a mixed integer linear program (MILP):
\begin{align}
    \begin{aligned}
        & \underset{\substack{\{x_{i,e}, b_{i}, f_{i,e}\}_{i\in\Ncal_{s},e\in\Ecal}}}{\text{minimize}}~~~
        && J~\text{in Eq.~(\ref{eq:cost})} \nn \\
        & \text{~~~~~~subject~to}
        && \text{Constraints in Eq.~(\ref{eq:slice_b_equal_sum_f}) and Table~\ref{table:constraints}}
    \end{aligned}
\end{align}
By solving this MILP, we determine the flow allocation and end-to-end routing path for each NTN slice.
The resources allocated to each gNB are subsequently distributed to its associated 5G UEs according to the policies defined by the 5G system~\cite{Abe_IAC22}.

\section{Numerical Simulations}

In this section, we verify the effectiveness of the proposed framework by focusing on the 5QI–NQI mapping function and its impact on NTN performance across different schemes.
We also compare the effect of the cost function weights in Eq.~(\ref{eq:cost}), which determine the priority between flow allocation and latency.

\subsection{Simulation Setup and Evaluation Metrics}

\begin{table}[tb]
    \centering
    \caption{Simulation parameters for users, traffic, satellites, and ground stations.}
    \scalebox{1}{
        \begin{tabular}{ll} \hline
            Parameter & Value \\ \hline \hline
            \textbf{User and traffic} \\
            Number of gNBs as NTN users & 30 \\
            Number of 5G UE & 150 (five per gNB) \\
            Number of traffic from each 5G UE & 20, 40, 60, 80, 100 \\
            \hline
            \textbf{LEO satellite} \\
            Number of LEO satellites & 90 \\
            ~~Number of planes & 6 \\
            ~~Number of satellites in each plane & 15 \\
            Altitude & 1,000~km \\
            Inclination & 90~deg \\
            \hline
            \textbf{Optical ground station} \\
            Number of ground stations & 3 \\
            \hline
            \textbf{Capacity} \\
            User link & 500~Mbps \\
            Inter-satellite link & 10~Gbps \\
            Feeder link & 10~Gbps \\
            \hline
            \textbf{Optimization} & \\
            Weights in cost function $(w_{f},w_{\ell})$ & (0.5, 0.5), (0.3, 0.7) \\
            \hline
        \end{tabular}
    }
\label{table:simulation_parameters}
\end{table}

\begin{figure}[tb]
	\centering
	\includegraphics[width=0.95\columnwidth]{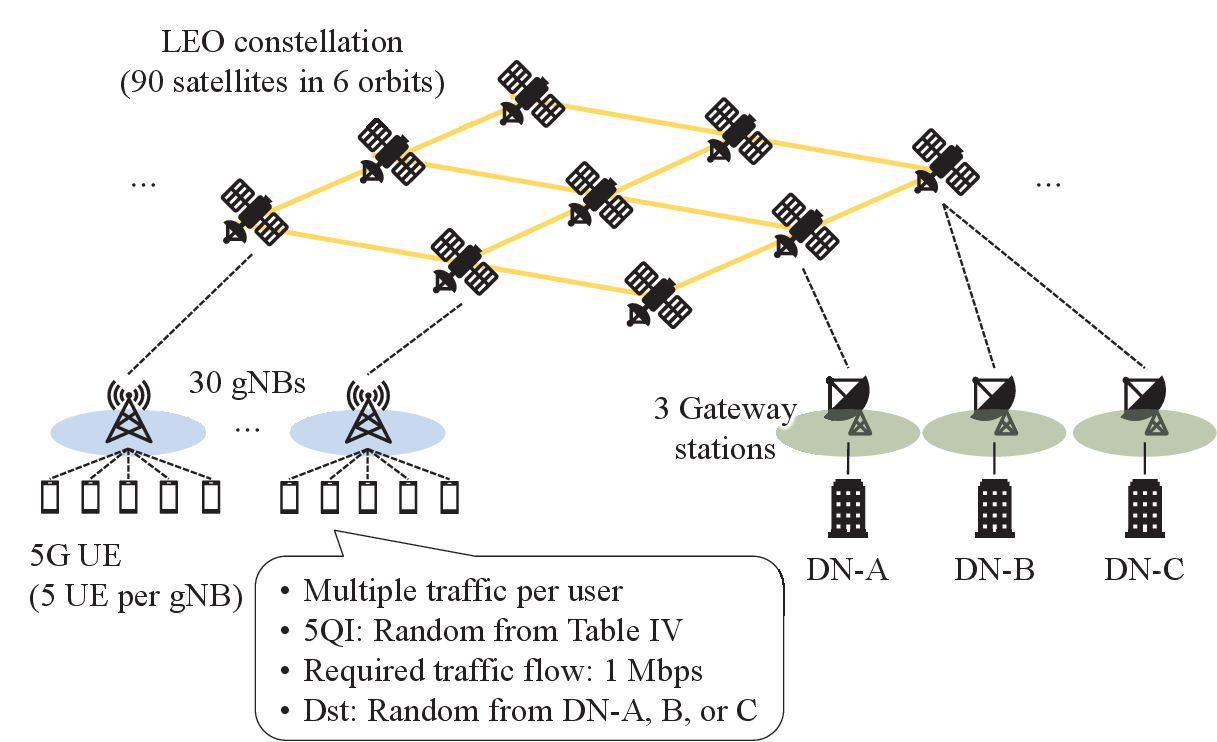}
	\caption{User, traffic, and network configuration in the simulation.}
\label{fig:simulation_configuration}
\end{figure}

Table~\ref{table:simulation_parameters} summarizes the simulation parameters, and Fig.~\ref{fig:simulation_configuration} illustrates the user, traffic, and network configuration.
LEO satellites are equipped with one radio frequency (RF) interface for user access and five optical interfaces: four for neighboring LEO satellites and one for an optical ground station (OGS).
The capacity of optical inter-satellite and feeder links is 10~Gbps, while RF user links support 500~Mbps.
Three OGSs are deployed in Tokyo (35.71$^{\circ}$, 139.49$^{\circ}$), western North America (42.45$^{\circ}$, –117.62$^{\circ}$), and southern Africa (–22.56$^{\circ}$, 19.03$^{\circ}$); the latter two correspond to candidate sites A and F in~\cite{Baeza_IEEENetwork24}.
Each OGS has one optical interface for LEO communication and is directly connected to a data network (DN), which serves as the destination for user traffic.

We generated 30 gNBs as NTN users, each directly connected to satellites via an RF interface and distributed globally.
Each gNB is connected to five 5G UEs, resulting in a total of 150 UEs.
Users generate multiple traffic flows, with the number ranging from 20 to 100 in steps of 20.
Each 5G traffic flow is assigned a destination uniformly at random from the three DNs defined above.
As described in Section~\ref{sec:nqi_slice_mapping}, the slice-edge satellite must be determined; here, we assume that each gNB connects to the nearest satellite.

This simulation uses the seven 5QIs listed in Table~\ref{table:5QI_simulation}, with the NQI set defined identically.
5QIs are assigned uniformly at random to each 5G traffic flow, with each flow requiring 1~Mbps.
The 5QI to NQI mapping function $f$ is defined in Table~\ref{table:5QI_NQI_mapping}.
Conditions~1 and 2 map all 5QIs to a single NQI (1 and 4), imposing strict (100~ms) and relaxed (300~ms) latency constraints, respectively.
Conditions~3 and 4 map the seven 5QIs to three and five NQIs, assigning each to the NQI with the largest PDB among grouped 5QIs.
Conditions~5 and 6 map the seven 5QIs one-to-one to seven NQIs: Condition 5 follows index order, while Condition 6 reverses it based on PDB values.
An NTN slice is defined as a set of NTN traffic flows with the same NQI and destination DN, as determined by the mapping function at the slice-edge satellites.
All gNBs and satellites are assumed to use common mapping functions $f$ and $g$, respectively.

The cost function weights $(w_{f},w_{\ell})$ are set in two configurations, as shown in Table~\ref{table:simulation_parameters}, and results are compared accordingly.
The normalization parameters $F$ and $L$ in Eq.~(\ref{eq:cost}) are defined as the total required flow and total PDB of all 5G traffic, respectively.
The flow allocated to each NTN slice by solving the optimization problem described in Section~\ref{sec:slice_optimization} is first distributed among its associated NTN users (a subset of gNBs) in proportion to their requested traffic and then further divided among each gNB’s 5G UEs proportionally.

\begin{table}[t]
\vspace*{0.2cm}
  \centering
  \caption{5QI values used in the simulation, with their standardized PDB according to~\cite{3GPP_TS23.501}. Values are sorted by ascending PDB. The same values are used to define NQIs.}
    \scalebox{1}{
    \begin{tabular}{c|ccccccc} \hline 
    5QI Value & 80 & 3 & 65 & 1 & 2 & 70 & 4 \\ \hline
    PDB [ms] & 10 & 50 & 75 & 100 & 150 & 200 & 300 \\
    \hline
    \end{tabular}
    }
\label{table:5QI_simulation}
\end{table}

\begin{table}[tb]
    \centering
    \caption{5QI-to-NQI mapping configurations used in the simulation, 
    expressed as $q_{n}=f(q_{f})$ under different conditions.}
    \begin{tabular}{ccl}
        \hline
        Condition & \#NQIs & Mapping $q_{n} \leftarrow q_{f}$ \\
        \hline \hline
        1 & 1 & 1 $\leftarrow$ [80, 3, 65, 1, 2, 70, 4] \\
        2 & 1 & 4 $\leftarrow$ [80, 3, 65, 1, 2, 70, 4] \\
        3 & 3 & 65 $\leftarrow$ [80, 3, 65]; 2 $\leftarrow$ [1, 2]; 4 $\leftarrow$ [70, 4] \\
        4 & 5 & \begin{tabular}[t]{@{}l@{}}%
                80 $\leftarrow$ 80; 65 $\leftarrow$ [3, 65]; 1 $\leftarrow$ 1; \\
                70 $\leftarrow$ [2, 70]; 4 $\leftarrow$ 4
            \end{tabular} \\
        5 & 7 & \begin{tabular}[t]{@{}l@{}}%
                80 $\leftarrow$ 80; 3 $\leftarrow$ 3; 65 $\leftarrow$ 65; \\
                1 $\leftarrow$ 1; 2 $\leftarrow$ 2; 70 $\leftarrow$ 70; 4 $\leftarrow$ 4
            \end{tabular} \\
        6 & 7 & \begin{tabular}[t]{@{}l@{}}%
                80 $\leftarrow$ 4; 3 $\leftarrow$ 70; 65 $\leftarrow$ 2; \\
                1 $\leftarrow$ 1; 2 $\leftarrow$ 65; 70 $\leftarrow$ 3; 4 $\leftarrow$ 80
            \end{tabular} \\
        \hline
    \end{tabular}
    \label{table:5QI_NQI_mapping}
\end{table}

To evaluate the simulation results, we define the user-averaged satisfaction ratios for flow allocation and latency as
\begin{align}
    \bar{s}_{f} = \dfrac{1}{\bar{N}_{u}}\sum_{i\in\bar{\Ncal}_{u}}\dfrac{b^{*}_{i}}{r_{f,i}}~~\text{and}~~\bar{s}_{\ell} = \dfrac{1}{\bar{N}_{u}}\sum_{i\in\bar{\Ncal}_{u}}\left(1-\dfrac{\ell^{*}_{i}}{\bar{\ell}(q_{f,i})}\right), \nn
\end{align}
where $b^{*}_{i}$ and $\ell^{*}_{i}$ are the allocated flow and resulting latency for the $i$-th traffic flow, respectively.
$\bar{N}_{u}=\sum_{j\in\Ncal_{g}}N_{u,j}$ and $\bar{\Ncal}_{u}$ denote the total number and the set of 5G traffic flows.
Higher values indicate better performance; however, $\bar{s}_{\ell}$ may be negative when $\ell^{*}_{i}>\bar{\ell}(q_{f, i})$ on average.

\subsection{Simulation Results}

\begin{figure*}[tb]
    \centering
    \begin{minipage}[t]{0.48\hsize}
        \centering
        \subfloat[Result of $\bar{s}_{f}$ with $(w_{f},w_{\ell})=(0.5,0.5)$.]{
            \includegraphics[width=0.87\columnwidth]{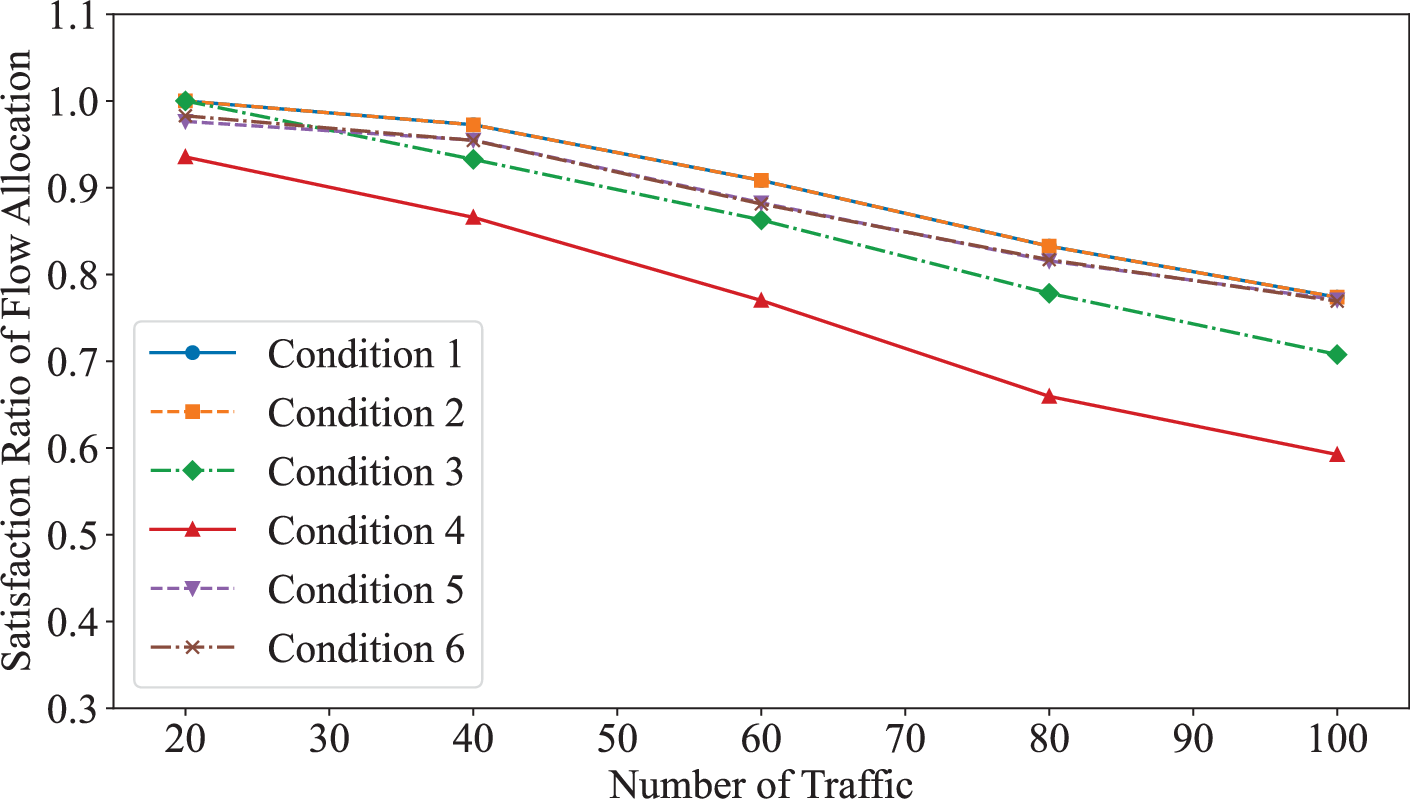}
            \label{fig:result_satisfaction_flow_05_05}
        }
    \end{minipage}
    \hfill
    \setcounter{subfigure}{2} 
    \begin{minipage}[t]{0.48\hsize}
        \centering
        \subfloat[Result of $\bar{s}_{f}$ with $(w_{f},w_{\ell})=(0.3,0.7)$.]{
            \includegraphics[width=0.87\columnwidth]{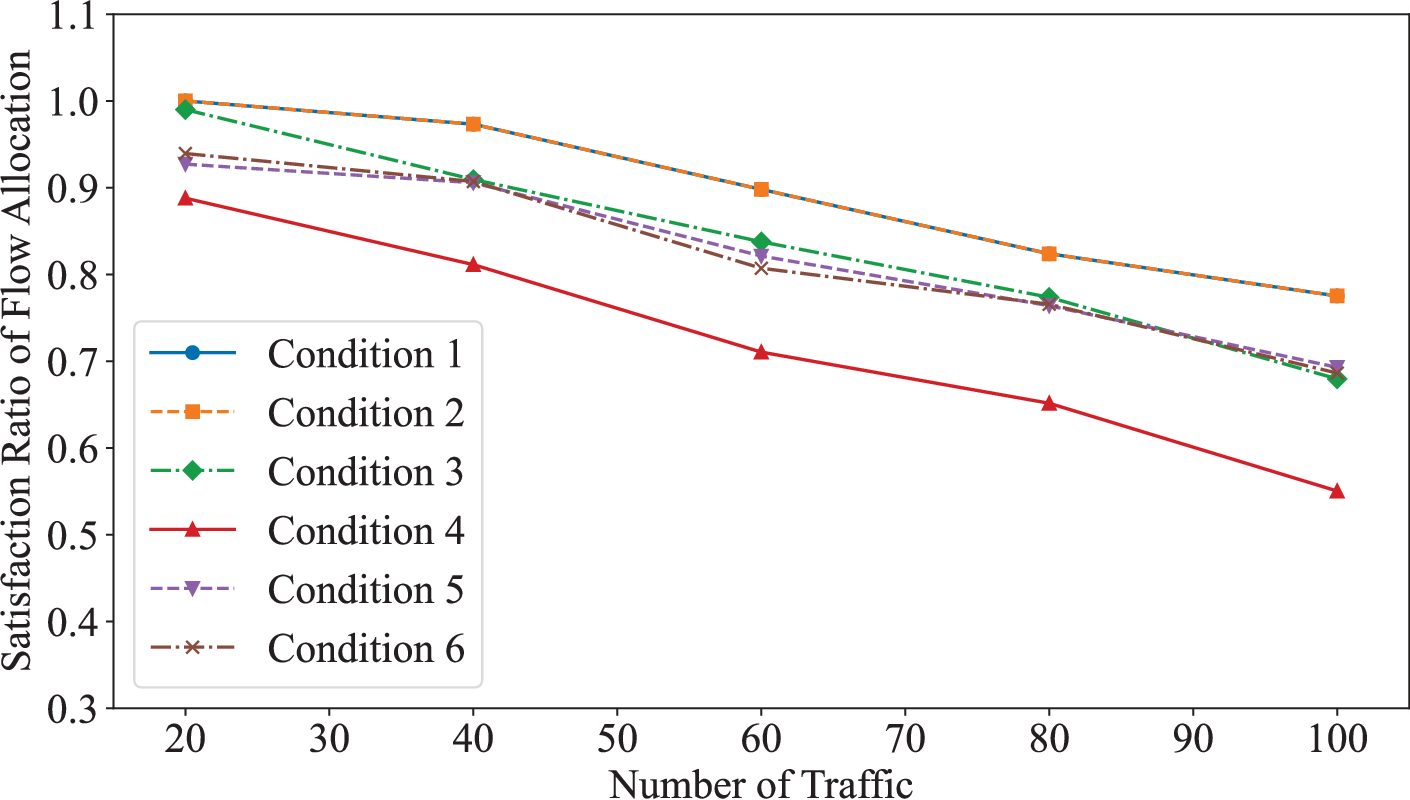}
            \label{fig:result_satisfaction_flow_03_07}
        }
    \end{minipage}
    \\ \medskip
    \setcounter{subfigure}{1} 
    \begin{minipage}[t]{0.48\hsize}
        \centering
        \subfloat[Result of $\bar{s}_{\ell}$ with $(w_{f},w_{\ell})=(0.5,0.5)$.]{
            \includegraphics[width=0.95\columnwidth]{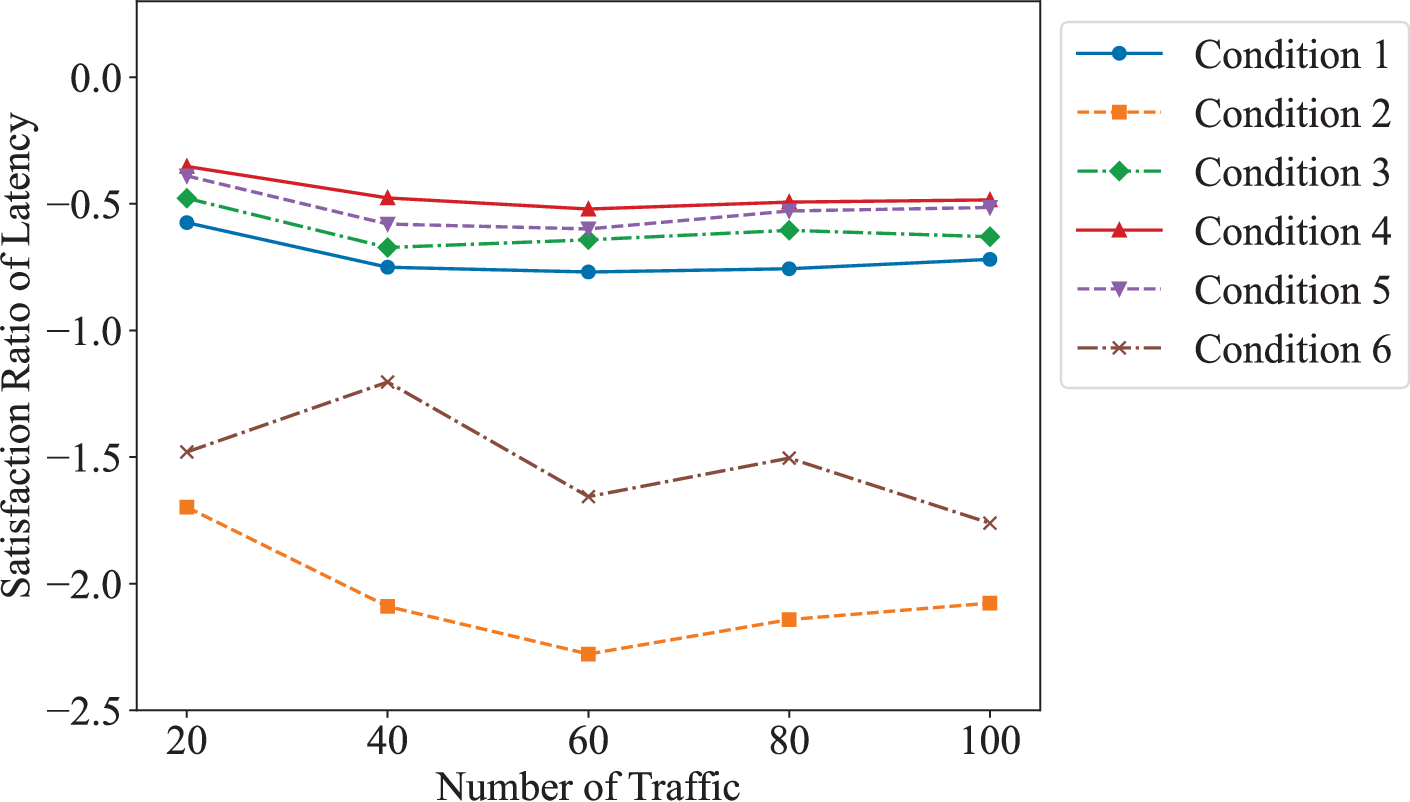}
            \label{fig:result_satisfaction_latency_05_05}
        }
    \end{minipage}
    \hfill
    \setcounter{subfigure}{3} 
    \begin{minipage}[t]{0.48\hsize}
        \centering
        \subfloat[Result of $\bar{s}_{\ell}$ with $(w_{f},w_{\ell})=(0.3,0.7)$.]{
            \includegraphics[width=0.95\columnwidth]{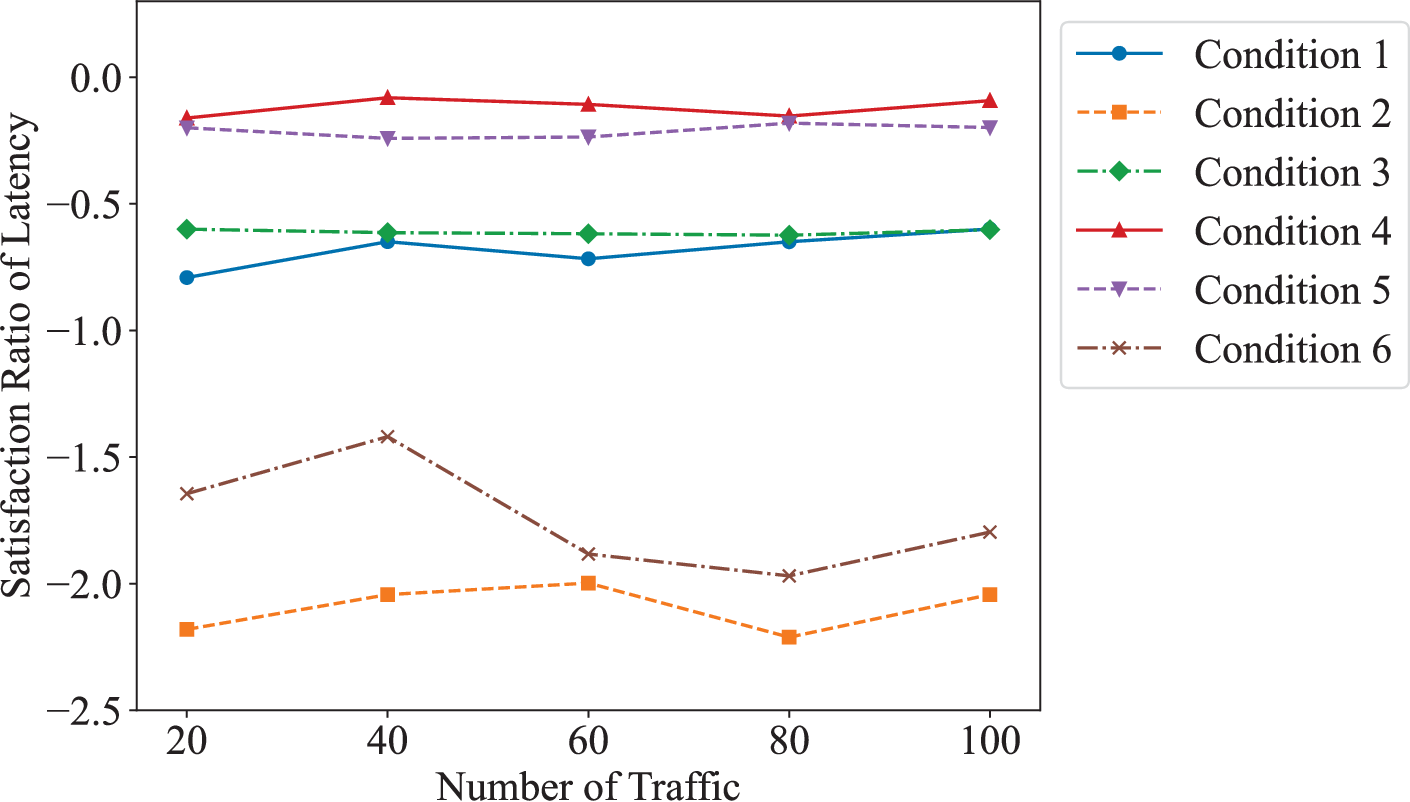}
            \label{fig:result_satisfaction_latency_03_07}
        }
    \end{minipage}
    \caption{Results of the user-averaged satisfaction ratio of the flow $\bar{s}_{f}$ and latency $\bar{s}_{\ell}$ with the different conditions of the mapping (Conditions~1 to 6) and weights in the cost function.}
    \label{fig:results_all_conditions}
\end{figure*}

Fig.~\ref{fig:results_all_conditions} compares user-averaged satisfaction ratios for flow and latency under different 5QI–NQI mapping conditions and cost function weights.
In Fig.~\ref{fig:result_satisfaction_flow_05_05}, flow satisfaction decreases with increasing traffic due to limited system capacity.
Among the six conditions, Condition~4 performs slightly worse, while others show minimal differences.
Fig.~\ref{fig:result_satisfaction_latency_05_05} shows that latency satisfaction is generally negative, indicating that NTN latency often exceeds the PDB required by 5G traffic.
Condition~2, which maps all 5QIs to NQI 4 (highest PDB), shows the worst result, followed by Condition~6, which reverses the PDB order.
The remaining four conditions yield satisfaction values around –0.5, corresponding to a latency ratio of $\ell^{*}_{i}/\bar{\ell}(q_{f, i})=1.5$, indicating that the assigned latency is roughly 1.5 times the required PDB on average.
Among these, Conditions~4 and 5, which preserve the PDB order with an equal or slightly reduced number of NQIs, show better performance.

As shown in Figs.~\ref{fig:result_satisfaction_flow_05_05} and \ref{fig:result_satisfaction_flow_03_07}, flow satisfaction slightly decreases when its weight in the cost function is reduced.
In contrast, Figs.~\ref{fig:result_satisfaction_latency_05_05} and \ref{fig:result_satisfaction_latency_03_07} show that latency satisfaction for Conditions~4 and 5, which already performed well with the weight of (0.5, 0.5), further improves as the latency weight increases.
These results highlight the impact of cost function weights on optimization performance.

\begin{figure}[tb]
	\centering
	\includegraphics[width=0.85\columnwidth]{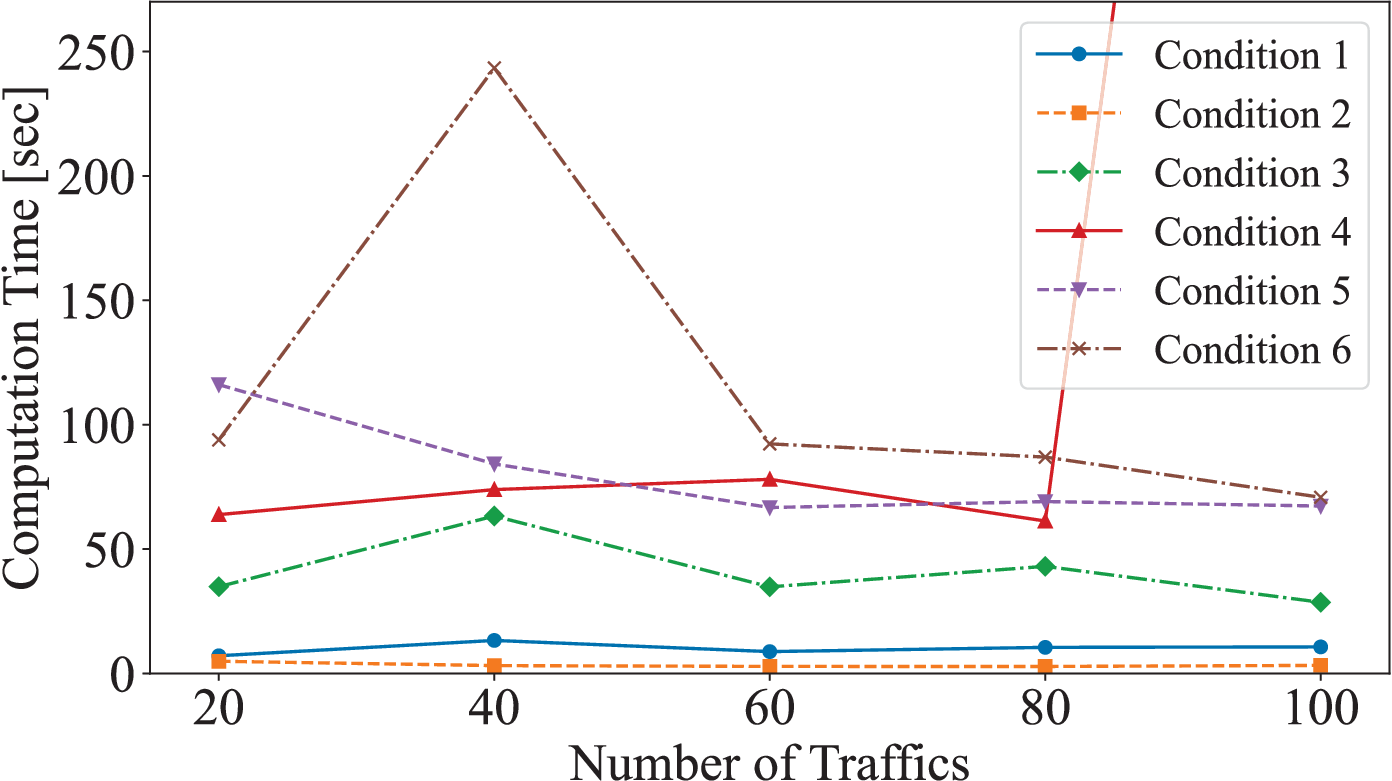}
	\caption{Result of the calculation time required for solving the optimization problem with $(w_{f},w_{\ell})=(0.5,0.5)$.}
\label{fig:result_calculation_time_05_05}
\end{figure}

Fig.~\ref{fig:result_calculation_time_05_05} shows the computation time for each parameter setting under $(w_{f},w_{\ell})=(0.5, 0.5)$.
All optimization problems were solved using the Gurobi Optimizer on an Intel Xeon w5-3425 CPU (3.19~GHz, 128~GB RAM).
Condition~4 with 100 traffic flows required 891.69~seconds, significantly higher than others, and is excluded from the plot.
Conditions with fewer NQIs in the 5QI–NQI mappings (Table~\ref{table:5QI_NQI_mapping}), such as Conditions~1 and 2, result in shorter computation times, while those with more NQIs (Conditions~4–6) tend to require longer.
This indicates a trade-off between latency satisfaction (as seen in Fig.~\ref{fig:result_satisfaction_latency_05_05}) and computation time, particularly for Conditions~4 and 5.

\section{Conclusion}

In this paper, we proposed the optimization framework for end-to-end traffic flow in 5G-NTN interconnected systems, consisting of (i) 5QI-to-NQI mapping, (ii) NTN traffic-to-slice mapping, and (iii) slice-level flow and routing optimization.
We evaluated several 5QI-to-NQI mapping schemes and found that the choice of mapping strongly affects flow allocation, latency satisfaction, and computation time.
Mappings that preserve the original PDB order and use a moderate number of NQIs (e.g., Conditions~4 and 5) yield higher latency satisfaction but require longer computation, highlighting a trade-off between performance and efficiency.

In future work, we will investigate scenarios including non-5G traffic and aim to optimize the mapping functions to maximize the performance of the proposed framework.
We also plan to incorporate fairness metrics to evaluate the framework, particularly under high-load scenarios.

\section*{Acknowledgment}

This study is conducted under the commissioned research of the ``Research and Development of Ka-band Satellite Communication Control for Various Use Cases'' (JPJ000254) by the Ministry of Internal Affairs and Communications, Japan.

\bibliographystyle{IEEEtran}
\bibliography{myref.bib}

\end{document}